\documentclass[conference]{IEEEtran}
\usepackage{graphicx}
\usepackage{amsfonts}
\usepackage[cmex10]{amsmath}
\usepackage{cite}
\usepackage{amssymb}
\usepackage{algorithmic}
\usepackage[center]{caption}
\usepackage{epsfig}
\usepackage{latexsym}
\usepackage{epstopdf}
\usepackage{amsmath, amsthm, amssymb}
\usepackage{verbatim}
\usepackage{amsthm}
\usepackage[linesnumbered,ruled,vlined]{algorithm2e}

\graphicspath{{./figures/}}
\usepackage{subfig}
\usepackage[usenames, dvipsnames]{color}
\usepackage{soul}
\definecolor{ultramarine}{rgb}{0.07, 0.04, 0.56} 
\begin{document}

\title{The Capacity Region of the Deterministic Y-Channel with Common and Private Messages}

\author{\IEEEauthorblockN{Mohamed S. Ibrahim \IEEEauthorrefmark{3},
		Mohammed Nafie\IEEEauthorrefmark{2} and
		Yahya Mohasseb\IEEEauthorrefmark{2}\IEEEauthorrefmark{1}		
	}
	\IEEEauthorblockA{	\IEEEauthorrefmark{3}Department of Electrical and Computer Engineering, University of Minnesota, Minneapolis, USA\\
		\IEEEauthorrefmark{2}Wireless Intelligent Networks Center (WINC), Nile University, Cairo, Egypt\\
		\IEEEauthorrefmark{1}Department of Communications, The Military Technical College, Cairo, Egypt\\
		Email: \{\tt youss037@umn.edu, mnafie@nileuniversity.edu.eg, mohasseb@ieee.org\}
	}
}
\maketitle
\begin{abstract}
In this paper, we discuss the capacity region of the deterministic 
Y-channel with private and common messages. 
Each user aims to deliver two private messages to
the other two users in addition to a common message directed towards both of them. As there is no direct link between the users, all messages must pass through the relay. We present outer-bounds on the rate region using genie aided and cut-set bounds. Then, we define an achievable region and show that at a certain number of levels at the relay, 
our achievable region coincides with the upper bound.
Finally, we argue that these bounds for this setup are not sufficient to characterize the capacity region.
\end{abstract}

\IEEEpeerreviewmaketitle
\section{Introduction}
Network information theory is one of the major challenges in wireless 
communication, which produced limited excellent results 
\cite{cover2012elements}. It involves the fundamental limits of 
communication and information theory in networks with multiple senders, 
receivers, optimal coding techniques and protocols which achieve these limits. 
An important goal is to characterize the capacity 
region or optimal rate which is the set of rates in which there exist codes with 
reliable transmissions. These rates of tuples are known to be achievable. A lot of 
work was done to determine the capacities of Gaussian networks with multiple 
senders and receivers. 

A deterministic model of the wireless channel -which is linear, 
and easy to analyze- was introduced in \cite{avestimehr2011wireless}, 
where the capacity of linear deterministic networks 
with a single source-destination pair is determined. Most subsequent research on
the linear deterministic model has been focused on deriving network coding schemes to 
determine the capacity of deterministic networks and then using the results 
to find an approximate capacity region for each corresponding Gaussian network,
where the approximation error can be typically ignored in the high Signal-to-Noise 
Ratio (SNR) regime. 
Using this model, the deterministic capacity region of different networks with only private
messages has been characterized in \cite{avestimehr2009capacity, mokhtar2010deterministic,
	chaaban2011capacity,sezgin2009approximate,zewail2013deterministic,
	chaaban2013approximate}.   

The authors of \cite{avestimehr2009capacity} characterized the 
deterministic capacity region of the multi-pair bidirectional 
wireless relay network and then, using the gleaned insights, 
proposed a transmission strategy for the gaussian two-pair two-way 
full duplex relay network and found an approximate characterization 
of the capacity region \cite{sezgin2009approximate}.
The authors of \cite{mokhtar2010deterministic} developed a new tighter upper 
bound based on the notion of one-sided genie. They utilized this 
bound to completely characterize the multicast deterministic capacity 
of a two pair symmetric bidirectional half duplex wireless relay network 
with only private messages. 
The authors of \cite{chaaban2011capacity} 
utilize the genie-aided bound developed in \cite{mokhtar2010deterministic} to 
characterize the capacity region of the deterministic Y channel with private 
messages only. They established the achievability by using three schemes: 
bidirectional, cyclic, and uni-directional communication. 
In \cite{chaaban2013approximate}, the approximate capacity 
of the gaussian Y-channel was obtained.
The authors of \cite{zewail2013deterministic}, 
characterize the deterministic capacity region of a four-node relay network 
with no direct links between the nodes.

In this paper, we study the deterministic capacity region of the Y-channel 
with private and common messages. This work is the first to deal with common messages, 
as all the previous work stated above deals only with private messages.
The Y-channel consists of three users and one intermediate 
relay, where each user aims to convey two private messages to the other two users 
in addition to a common message to both of them via the intermediate relay. 
First, we use cut-set bound and genie-aided bounds to obtain an outer bound on the deterministic capacity region of the network. 
Then, we define an achievable rate region of the network and show that it coincides 
with the upper bound at a certain number of levels at the users.
Finally, we develop a greedy strategy, namely the Gain Ordering Scheme (GOS), 
to send messages over the network. This strategy is then used to characterize 
the achievable rate region of the network. In principle, the GOS starts by sending 
the bits which can be combined effectively at the relay and ends with the ones that 
must be sent individually.

The rest of the paper is organized as follows. We describe 
the system model and state our main results in section 
\ref{system_model}. In Section \ref{ACHIEVABILITY}, we detail 
our achievability schemes. In Section \ref{Upper_bound_common}, we discuss 
the genie-aided bounds in the existence of common messages. In Section 
\ref{disc}, we mention some insights about our work and illustrate our 
achievability scheme with two numerical examples. Finally, Section 
\ref{conc} presents our conclusions.  

\begin{figure}[t]
	\centering
	\includegraphics[width=0.88\linewidth]{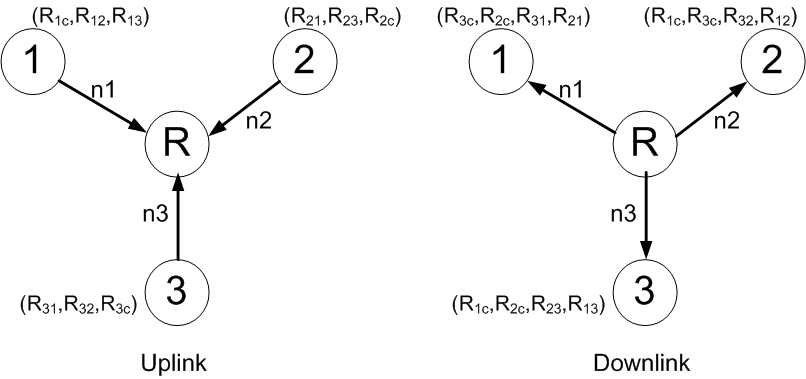}
	\caption{System Model}
	\label{sys_model}
\end{figure}

\section{System Model}
\label{system_model} 
\begin{figure}[t]
	\centering
	\includegraphics[width=0.65\linewidth]{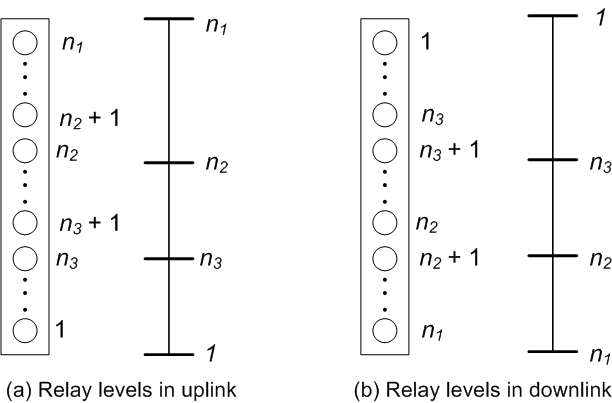}
	\caption{ Relay levels enumeration    \label{enumeration}}
\end{figure}

Consider the multi-way relaying network, shown in Fig. \ref{sys_model},
where user $j$ aims to convey two private messages 
with rates $R_{ji}$ and $R_{jk}$ to users  $i$ and $k$, respectively, 
in addition to a common message to both of them with rate $R_{jc}$ via the intermediate relay, 
for $i,j,k \in \{1,2,3\}$ and $i \neq j \neq k$. 
Due to the absence of direct links between the users, 
communication occurs through the relay in two phases: 
uplink phase and downlink phase. 
All nodes are assumed to be full duplex (i.e., the nodes can transmit and 
receive simultaneously).
In a deterministic channel model \cite{avestimehr2011wireless}, the 
channel can be defined as the number of levels between 
each user and the relay, where each level can be represented as a circle as Fig \ref{enumeration} depicts. 
We denote the channel between user $j$ and the relay by $n_j$, 
where $j \in \{1,2,3\}$. We assume that the channel is reciprocal,
i.e., ($n_{jR} = n_{Rj} = n_{j}$) where $ n_j = \lceil 0.5\log _2{\rho} \rceil $ and 
$\rho$ is the SNR. 
We can assume without loss of generality that
\begin{equation}
n_1 \geq n_2 \geq n_3
\end{equation}
otherwise we can re-label the nodes. \\

An outer bound on the deterministic capacity region of this network is given by the following theorem:\\
\textbf{Theorem 1.} \textit{ The capacity region $\mathcal{C}$ of 
	the deterministic Y channel is given by the rates which satisfy the following conditions }
\begin{equation} \label{UB_1}
R_{1c} + R_{2c} + R_{13} + R_{23} \leq n_3
\end{equation}
\begin{equation}\label{UB_2}
R_{1c} + R_{12} + R_{13} + R_{32} + R_{3c} \leq n_2
\end{equation}
\begin{equation}\label{UB_3}
R_{1c} + R_{12} + R_{13} +  R_{23} + R_{2c} \leq n_2
\end{equation}
\begin{equation}\label{UB_4}
R_{3c} + R_{31} + R_{32} \leq n_3
\end{equation}
\begin{equation}\label{UB_5}
R_{2c} + R_{21} + R_{3c} + R_{31} +  R_{32} \leq n_2
\end{equation}
\begin{equation}\label{UB_6}
R_{2c} + R_{21} + R_{3c} + R_{31} + R_{23}  \leq n_2
\end{equation}
\begin{equation} \label{Ub_7}
R_{2c} + R_{21} + R_{23} + R_{13} + R_{1c} \leq n_1  
\end{equation}
\begin{equation} \label{Ub_8}
R_{3c} + R_{31} + R_{32} + R_{12} + R_{1c} \leq n_1  
\end{equation}
where $R_{ij}$ is the rate from node $i$ to node $j$ and $R_{ic}$ is the rate from 
node $i$ to both nodes $j$ and $k$, where 
$ i,j,k\in{\left\lbrace1,2,3\right\rbrace }$ and $i \neq j \neq k $. 
This capacity region is outer bounded by $\mathcal{\bar{C}}$, where 
\begin{align}
\mathcal{\bar{C}}\, \triangleq\, \big\lbrace\  \text{\textbf{R}} \subset \mathbb{R}_{+}^{9} \vert\quad \text{\textbf{R}} \,\ \text{satisfies}\,  (2)-(9)\ \big\rbrace 
\end{align}
The following Lemma represents the main result of the paper, \\
\textbf{Lemma 1.} \textit{The Gain Ordering Scheme achieves 
all the integral rate tuples in the intersection between the capacity 
region $\mathcal{C}$  stated in Theorem 1 and the following extra condition:}\\
\begin{equation} \label{L_con}
\min\left\{\begin{array}{l}
R_{3c} + R_{31} + R_{32} + R_{12} + R_{1c}\\
R_{2c} + R_{21} + R_{23} + R_{13} + R_{1c}      
\end{array} 
\right\} \leq n_2 
\end{equation}

The proof of Lemma 1 will be shown at the end of Section \ref{ACHIEVABILITY}. It is worth-mentioning that, when $n_1 = n_2$, our achievable region coincides with the upper bound.
This can be simply concluded by observing that the extra condition defined in Lemma 1 is equivalent to the upper bound equations (\ref{Ub_7}) and (\ref{Ub_8}) in Theorem 1.

\section{ACHIEVABILITY}
\label{ACHIEVABILITY}
In this section, we use the GOS 
to characterize the regions that can be achieved. The rationale of the GOS idea is to start by selecting the bits from the rate tuple $\textbf{R}$
that can together attain the maximum gain, as defined below. Then we eliminate the transmitted bits 
from $\textbf{R}$ and continue with the transmission of the remaining rate tuple using the intermediate gain strategy. 
Finally, after sending the bits which achieve the intermediate gain, we go to the minimum 
gain strategy with the residual rate tuple. 
Note that the relay does not need to decode the data and therefore, combining bits on the same level at the relay provides gains in achievable rates. However, as we shall see below, these gains differ according to the number of levels combined. 

\subsection{Maximum Gain Strategy}
This case occurs when two users want 
to exchange private bits. That is, $R_{ji}$ and $R_{ij}$ are both 
non-zero for some $j,i \in {\left\lbrace1,2,3\right\rbrace }$, $j \neq i$. 
Thus, sending two bits on the same level from the two users yields 
a gain of 2 bits/level. In particular, each user can decode its desired bit
by x-oring its own bit with the received signal. 
This scheme continues until we send $\min(R_{ij}, R_{ji})$ $ \forall$ $ i,j \in \{1,2,3\}$. Consider a rate tuple \textbf{R} $ \in \mathbb{R}_{+}^{9}$, where 
\begin{equation}\label{Max0}
\text{\textbf{R}} = \{ R_{12},R_{13},R_{1c},R_{21},R_{23},R_{2c},R_{31},R_{32},R_{3c}\} 
\end{equation}
Let

\begin{eqnarray} \label{Max1}
&a_{1}& = \min\left\lbrace R_{12},R_{21}\right\rbrace, 
a_{2} = \min\left\lbrace R_{13},R_{31}\right\rbrace, \nonumber \\ 
&a_{3}& = \min\left\lbrace R_{23},R_{32}\right\rbrace
\end{eqnarray}
In the uplink users 1 and 2 use levels $\left\lbrace 
n_{2} - a_{1} + 1,\cdots,n_{2}\right\rbrace$ to send their 
binary vectors $\boldsymbol{z}_{12}$ and $\boldsymbol{z}_{21}$. 
The relay obtains $\boldsymbol{z}_{12} \oplus \boldsymbol{z}_{21}$ 
and sends it back on the same levels. Now, user 1 and 2 -by knowing 
their information- can perfectly decode their desired information from 
$\boldsymbol{z}_{12} \oplus \boldsymbol{z}_{21}$. Similarly users 1 and 3 
use relay levels $\left\lbrace 1,\cdots,a_{2} \right\rbrace $. Also, users 
2 and 3 use levels $\left\lbrace a_{2}+1,\cdots,a_{2}+a_{3} \right\rbrace$.

To guarantee that this strategy works perfectly, we should have $a_{2}+a_{3} \leq n_{3}$ and $a_{1}+a_{2}+a_{3} \leq n_{2}$. Since $\text{\textbf{R}} \in \mathcal{\bar{C}}$ then \\
\begin{equation}\label{Max2}
    a_{2} + a_{3} \overset{(\ref{Max1})} \leq R_{31} + R_{32} \overset{(\ref{UB_4})} \leq n_{3} 
 \end{equation}    
\begin{equation}\label{Max3}
    a_{1} + a_{2} + a_{3} \overset{(\ref{Max1})} \leq R_{31} + R_{32} + R_{21} \overset{(\ref{UB_5})} \leq n_{2}
\end{equation}
and subsequently the levels of the relay are sufficient for this strategy 
to work. Now, we need to achieve
\begin{align}\label{Max4}
\text{\textbf{R}}_{I} = (&R_{12} - a_{1},R_{13} - a_{2},R_{1c},R_{21} - a_{1},R_{23} - a_{3}, \nonumber \\
 &R_{2c}, R_{31} - a_{2},R_{32} - a_{3},R_{3c})
\end{align}

Then, we remove the occupied relay levels, thus the available levels 
are now $n^I_{1}$, $n^I_{2}$ and $n^I_{3}$, where 
\begin{equation}\label{Max5}
 n^{I}_{1} = n_{1} - a_{1} - a_{2} - a_{3}
 \end{equation} 
\begin{equation}\label{Max6}
 n^{I}_{2} = n_{2} - a_{1} - a_{2} - a_{3}
\end{equation}
 Recall that in the communication between user 1 and 2, 
 we use the relay levels $\lbrace n_{2} - a_{1} + 1,\cdots,n_{2}\rbrace$. 
 This does not consume levels in $\lbrace 1,\cdots,n_{3}\rbrace$ if 
 $n_{2} - n_{3} \geq a_{1}$. In this case, $n^{I}_{3} = n_{3} - a_{2} - a_{3}$. 
 Otherwise, levels $a_{1} - (n_{2} - n_{3})$ are used in $1,\cdots,n_{3}$. 
 Subsequently, $n^{I}_{3} = n_{3} - a_{2} - a_{3} - (a_{1} - (n_{2} - n_{3})) 
 = n^{I}_{2}$. So,
\begin{equation}\label{Max7}
n^{I}_{3} = \min\lbrace n_{3} - a_{2} - a_{3},n^{I}_{2}\rbrace
\end{equation}  
The residual non zero rates in the vector \textbf{R} can be represented by any of the next strategies.
\subsection{Intermediate Gain Strategy}
We use this strategy when the minimum of 
$R_{ij}$ and $R_{ji}$ is equal to zero, $\forall$ $ i,j \in \{1,2,3\}$.
In this strategy, we start by the new rate tuple $\text{\textbf{R}}_I$ in (\ref{Max4}). 
The gain achieved from this strategy is 1.5 bits/level. 
There are four  different scenarios that can attain this gain. 
Since the common bit is intended to two users, so it is reasonable to give 
it a higher priority than the private one. As a result, we order the four scenarios 
according to the number of common bits relative to the number of private bits.
In what follows, we detail the four scenarios for achieving 1.5 bits/level as follows,
\begin{enumerate}
\item Full Common Scenario \\
In this case, each user has at least one common message 
i.e., ($R_{jc} \neq 0, \ j \in \lbrace 1,2,3\rbrace$). Let\\
\begin{equation} \label{Fcomm1}
b = \min\{ R_{1c},R_{2c},R_{3c} \}
\end{equation}
Let the transmit binary vectors of users 1, 2, and 3 be 
$\boldsymbol{z}_{1c}$, $\boldsymbol{z}_{2c}$, and $\boldsymbol{z}_{3c}$.
Users 1 and 2 send $\boldsymbol{z}_{1c}$ and $\boldsymbol{z}_{2c}$ 
on relay levels $\left\lbrace n^{I}_{2} - b + 1,\cdots,n^{I}_{2}\right\rbrace$. 
User 2 also repeats $\boldsymbol{z}_{2c}$ on relay levels $\lbrace 1,\cdots,b\rbrace$ 
together with user 3 which sends $\boldsymbol{z}_{3c}$ on the same levels. 
The relay receives $\boldsymbol{z}_{1c} \oplus \boldsymbol{z}_{2c}$ and 
$\boldsymbol{z}_{2c} \oplus \boldsymbol{z}_{3c}$ and then sends them back 
to all the users on the levels $\lbrace 1,\cdots,2b\rbrace$. 
All Users receive $\boldsymbol{m}_{1} = \boldsymbol{z}_{1c} \oplus \boldsymbol{z}_{2c}$ 
and $\boldsymbol{m}_{2} = \boldsymbol{z}_{3c} \oplus \boldsymbol{z}_{2c}$. 
Each user now can decode its intended messages without error. 
User 2 can decode $\boldsymbol{z}_{1c}$ and $\boldsymbol{z}_{3c}$ by x-oring its own bit
$\boldsymbol{z}_{2c}$ with
$\boldsymbol{m}_{1}$ and $\boldsymbol{m}_{2}$, respectively. 
Then, by x-oring $\boldsymbol{z}_{1c}$ with $\boldsymbol{m}_{1}$, user 1 can extract $\boldsymbol{z}_{2c}$ and afterwards, user 1 x-or $\boldsymbol{z}_{2c}$ with $\boldsymbol{m}_{2}$ to get 
$\boldsymbol{z}_{3c}$. Similarly, user 3 can decode $\boldsymbol{z}_{2c}$ and 
$\boldsymbol{z}_{1c}$ from $\boldsymbol{m}_{2}$ and $\boldsymbol{m}_{1}$, respectively. 
To guarantee the operation of this strategy, we need $2b \leq n^{I}_{3}$, (i.e., $ 2b + a_{2} + a_{3} \leq n_{3} $). Since
\begin{equation}\label{Fcomm3}
         2b + a_{2} + a_{3} \overset{(\ref{Max1}),(\ref{Fcomm1})} \leq R_{1c} + R_{2c} + R_{13} + R_{23} 
         \overset{(\ref{UB_1})} \leq n_{3}         
 \end{equation}              
Therefore, there are enough levels for serving all bits in this scenario. 
After removing all the occupied levels and subtracting the rates that are achieved. 
The rate tuple that still needs to be achieved is 
\begin{align}\label{Fcomm5}
 \textbf{R}_{I}^{'} = (&R_{12} - a_{1},R_{13} - a_{2},R_{1c}-b, R_{21} - a_{1}, \nonumber \\ 
 &R_{23} - a_{3},R_{2c}-b,R_{31} - a_{2},R_{32} - a_{3},R_{3c}-b) \nonumber \\  \triangleq (&R^{'}_{12`},R^{'}_{13},R^{'}_{1c},R^{'}_{21},R^{'}_{23},R^{'}_{2c},R^{'}_{31},R^{'}_{32},R^{'}_{3c})
\end{align}
and the new number of levels at each node will be as follows,
 \begin{equation}\label{Fcomm6}
 n^{'}_{1} = n^{I}_{1} - 2b
 \end{equation}
 \begin{equation}\label{Fcomm7}
 n^{'}_{2} = n^{I}_{2} - 2b
 \end{equation} 
 \begin{equation}\label{Fcomm8}
 n^{'}_{3} =  n^{I}_{3} - 2b
 \end{equation}


\item Two Common One Private Scenario \\
This scenario applies when one of the users has $R_{jc} = 0$ and 
$\min(R_{jk},R_{ji}) \neq 0$, for $i,j,k = \{1,2,3\}$ and $i \neq j \neq k$.
We aim to achieve $R^{'}_{I}$ with at least 3 zero private components and 1 zero common component.
Depending on which 3 private rates out of the 6 private rates in $R^{'}_{I}$ are zero, we have 8 different cases.
Similarly, depending on the zero common rate, we may have three different cases for zero common rate.    
However, as space limitations prohibit showing all the cases, we focus on one of them and the rest follows in the same way. Hence, we select the case where $R^{'}_{2c} = 0$ and 
$(R^{'}_{12},R^{'}_{23},R^{'}_{13}) = (0,0,0)$. Therefore, from (\ref{Fcomm5}), the rate vector that needs to be achieved is
\begin{equation}\label{TCOP1}
 \textbf{R}_\textbf{I}^{'} = (0,0,R^{'}_{1c},R^{'}_{21},0,0,R^{'}_{31},R^{'}_{32},R^{'}_{3c})
\end{equation}
Now, we start this scenario by defining the following
\begin{equation}\label{TCOP2}
  c_{1} = \min\lbrace R^{'}_{1c},R^{'}_{3c},R^{'}_{23} \rbrace
\end{equation}
\begin{equation}\label{TCOP3}
  c_{2} = \min\lbrace R^{'}_{1c} - c_{1} ,R^{'}_{3c} - c_{1},R^{'}_{21} \rbrace
\end{equation}
Let the transmit vectors of users 1, 2, and 3 be $\boldsymbol{z}_{1c}$, $\boldsymbol{z}_{2c}$, and $\boldsymbol{z}_{3c}$. Users 1 and 2 send $\boldsymbol{z}_{1c}$ and $\boldsymbol{z}_{21}$ on relay levels $\left\lbrace n^{'}_{2} - c + 1,\cdots,n^{'}_{2}\right\rbrace$. User 1 also repeats $\boldsymbol{z}_{1c}$ on relay levels $\lbrace 1,\cdots,b\rbrace$ together with user 3 which sends $\boldsymbol{z}_{3c}$ on the same levels. The relay receives $\boldsymbol{m}_{1} =\boldsymbol{z}_{1c} \oplus\boldsymbol{z}_{21}$ \text{and} 
$\boldsymbol{m}_{2} = \boldsymbol{z}_{1c} \oplus \boldsymbol{z}_{3c}$ and sends them back on the same levels. Users 1 and 2 receive $ \boldsymbol{m}_{1} $ and $ \boldsymbol{m}_{2} $. User 3 receives $\boldsymbol{m}_{2}$ only. User 1, knowing $\boldsymbol{z}_{1c}$, can easily extract $\boldsymbol{z}_{21}$ and $\boldsymbol{z}_{3c}$ from $\boldsymbol{m}_{1}$ and $\boldsymbol{m}_{2}$, respectively.Then, knowing $\boldsymbol{z}_{21}$, user 2 can decode $\boldsymbol{z}_{1c}$ from $\boldsymbol{m}_{1}$ and afterwards decodes $\boldsymbol{z}_{3c}$ from $\boldsymbol{m}_{2}$. Finally, user 3 extracts $\boldsymbol{z}_{1c}$ from $\boldsymbol{m}_{2}$. 
This strategy works if we have enough levels at the relay that allow the transmission and reception of 
$ c_1 \space$ and $\space c_2 $, i.e., $2c_{1}+c_2 \leq n^{'}_{3}$ , $ 2c_{1}+2c_2 \leq n^{'}_{3}$.

\begin{align}\label{TCOP4}
 2c_{1}+c_{2} & \overset{(\ref{TCOP3})}\leq 2c_{1} + R^{'}_{1c} - c_{1} = R^{'}_{1c} + c_{1} \\ 
    & \overset{(\ref{TCOP4})}\leq R^{'}_{1c} + R^{'}_{23}  \overset{(\ref{Fcomm5})}= R_{1c} - b + R_{23} - a_{3} \\
    & \overset{(\ref{UB_1})}\leq n_{3} - R_{2c} - b - R_{13} - a_{3} \\ 
     &\overset{(\ref{Max1}),(\ref{Fcomm1})}\leq n_{3} - 2b - a_{2} - a_{3} \overset{(\ref{Max7}),(\ref{Fcomm8})}= n^{'}_{3}    
\end{align}

\begin{align}\label{TCOP5}
 2c_{1}+2c_{2} & \overset{(\ref{TCOP3})}\leq 2c_{1} + R^{'}_{1c} + R^{'}_{3c} - 2c_{1}  \\ 
    &\overset{(\ref{Fcomm5})}= R_{1c} + R_{3c} - 2b \\ & \overset{(\ref{UB_2})}\leq n_{2} - R_{12} - R_{13} - R_{32} - 2b \\
    &\overset{(\ref{Max1})}= n_{2} - a_{1} - a_{2} - a_{3} - 2b \overset{(\ref{Max6}),(\ref{Fcomm7})}= n^{'}_{2}    
 \end{align} 
Now considering the case where $R^{'}_{3c}$ is the minimum in (\ref{TCOP2}). The remaining vector that needs to be achieved is
\begin{align}\label{TCOP6}
 \textbf{R}_\textbf{I}^{''} &= (0,0,R^{'}_{1c} - c_{2},R^{'}_{21}-c_{2},0,0,R^{'}_{31},R^{'}_{32},R^{'}_{3c} - c_{2}) \nonumber\\ &\triangleq (0,0,R^{''}_{1c},R^{''}_{21},,0,0,R^{''}_{31},R^{''}_{32},0)  
\end{align}   
 with number of levels at each node $(n^{''}_{1},n^{''}_{2},n^{''}_{3})$ where,
 \begin{equation}\label{TCOP7}
 n^{''}_{1} = n^{'}_{1} - 2c_{2}
 \end{equation}
 \begin{equation}\label{TCOP8}
 n^{''}_{2} = n^{'}_{2} - 2c_{2}
 \end{equation}
if $ c_{2} \leq n^{'}_{2} - n^{'}_{3} $ then $ n^{''}_{3} = n^{'}_{3} - c_{2} $, otherwise, $ n^{''}_{3} = n^{'}_{3} - c_{2} - ( c_{2} - (n^{'}_{2} -n^{'}_{3})) = n^{''}_{2}$. Therefore,

 \begin{equation}\label{TCOP9}
 n^{''}_{3} = \min\lbrace n^{'}_{3} -c_{2}, n^{''}_{2}\rbrace
 \end{equation}
\item One Common Two Private Scenario \\ 
For the operation of this scenario, there should be at least one of the users has non-zero common bits in addition to one of the following two conditions should be satisfied in the residual rate vector
\begin{itemize}
	\item Each of the other two users have a non-zero private bits to
	the user that has common bits.
    \item User $ j $ has two common bits to both users $ i,k $, user $ i $ has private bits to user $ j $ and user $ k $ has private bits to user $ i $, for $ i,j,k \in \{1,2,3\}$ and $   i \neq j \neq k $.
\end{itemize} 
\begin{align}
 d_{1} &= \min\lbrace R^{''}_{1c},R^{''}_{31},R^{''}_{23} \rbrace \label{OCTP1}, \\
 d_{2} &= \min\lbrace R^{''}_{1c},R^{''}_{21},R^{''}_{32} \rbrace \label{OCTP2} \\
 d_{3} &= \min\lbrace R^{''}_{1c} - d_{1} - d_{2},R^{''}_{31}-d_{1} ,R^{''}_{21}-d_{2} \rbrace \label{OCTP3}
\end{align}
Notice that, since we always start with the maximum gain phase, either $d_{1}$ or $d_{2}$ must be zero because 
one of $ R^{''}_{23}$ and $R^{''}_{23} $ should be zero. According to (\ref{TCOP1}), we should have $d_{1} $ equal zero. 
Let the transmit vectors of users 1, 2, and 3 be $\boldsymbol{z}_{1c}, \boldsymbol{z}_{21}, \boldsymbol{z}_{32}$ and $\boldsymbol{z}_{31}$. Users 1 and 2 send $\boldsymbol{z}_{1c}$ and $\boldsymbol{z}_{21}$ on relay levels $\left\lbrace n^{''}_{2} - d_{2} + 1,\cdots,n^{'}_{2}\right\rbrace$. User 1 also repeats $\boldsymbol{z}_{1c}$ on relay levels $\lbrace 1,\cdots,d_{2}\rbrace$ together with user 3 which sends $\boldsymbol{z}_{32}$ on the same levels. The relay receives $\boldsymbol{z}_{1c} \oplus \boldsymbol{z}_{21}$ and $\boldsymbol{z}_{1c} \oplus \boldsymbol{z}_{3c}$ and sends them back on the same levels. Users 1 and 2 receive $ \boldsymbol{w}_{1} = \boldsymbol{z}_{1c} \oplus \boldsymbol{z}_{21} $ and $ \boldsymbol{w}_{2} = \boldsymbol{z}_{1c} \oplus \boldsymbol{z}_{32} $. User 1, knowing $\boldsymbol{z}_{1c}$, can easily extract $\boldsymbol{z}_{21}$ and $\boldsymbol{z}_{32}$ from $\boldsymbol{w}_{1}$ and $\boldsymbol{w}_{2}$, respectively. Then, knowing $\boldsymbol{z}_{21}$, user 2 can decode $\boldsymbol{z}_{1c}$ from $\boldsymbol{w}_{1}$ and afterwards decodes $\boldsymbol{z}_{32}$ from $\boldsymbol{w}_{2}$. Finally, user 3 extracts $\boldsymbol{z}_{1c}$ from $\boldsymbol{w}_{2}$.

\par Now, we start transmitting $ d_3 $ bits, users 1 and 2 send $\boldsymbol{z}_{1c}$ and $\boldsymbol{z}_{21}$ on relay levels $\{n^{''}_{2} - d_{2} - d_{3} + 1,\cdots,n^{'}_{2}-d_{2}\}$. User 1 also repeats $\boldsymbol{z}_{1c}$ on relay levels $\lbrace d_{2}+1,....,d_{2}+d_{3}\rbrace$ together with user 3 which sends $\boldsymbol{z}_{31}$ on the same levels. The relay receives $\boldsymbol{z}_{1c} \oplus \boldsymbol{z}_{21}$ and $\boldsymbol{z}_{1c} \oplus \boldsymbol{z}_{31}$ and sends them back on the same levels. Users 1 and 2 receive $ \boldsymbol{w}_{1} = \boldsymbol{z}_{1c} \oplus \boldsymbol{z}_{21} $. User 1, knowing $\boldsymbol{z}_{1c}$, can easily extract $\boldsymbol{z}_{21}$ and $\boldsymbol{z}_{31}$ from $\boldsymbol{w}_{1}$ and $\boldsymbol{w}_{2}$, respectively. Then, knowing $\boldsymbol{z}_{21}$, user 2 can decode $\boldsymbol{z}_{1c}$ from $\boldsymbol{w}_{1}$. Finally, user 3 extracts $\boldsymbol{z}_{1c}$ from $\boldsymbol{w}_{2}$. This strategy works if we have enough levels at the relay, i.e., $2d_{2}+2d_{3} \leq n_{2}^{''}$, $d_{2}+d_{3} \leq n_{3}^{''}$, and by extracting $ n_{2}^{''} $ from equations (\ref{TCOP8}), (\ref{Fcomm8}) and (\ref{Max7}), and  $ n_{3}^{''} $ from equations (\ref{TCOP7}) , (\ref{Fcomm7}) and 
 (\ref{Max6}), respectively, we get the following equations after rearranging
 \begin{align}
 2d_{2}+2d_{3}+2c_{2}+2b+a_{1}+a_{2}+a_{3} &\leq n_{2} \label{OCTP4} \\
 d_{2}+d_{3}+c_{2}+2b+a_{2}+a_{3} &\leq n_{3} \label{OCTP5}
 \end{align}
In what follows, we prove equations (\ref{OCTP4}) and (\ref{OCTP5})
\begin{align*} 
 2d_{2} + 2d_{3} + 2c_{2} + 2b +a_{1}+a_{2}+a_{3} &\overset{(\ref{OCTP2}),(\ref{OCTP3})}\leq \nonumber \\
 R_{31}^{''} + R_{21}^{''} + R_{32}^{''} + 2c_{2} + 2b + a_{1}+a_{2}+a_{3} &\overset{(\ref{TCOP6}),(\ref{TCOP3})}\leq \nonumber\\
 R_{31}^{'} + R_{21}^{'} + R_{32}^{'} + R_{3c}^{'} + 2b + a_{1}+a_{2}+a_{3} &\overset{(\ref{Fcomm1}),(\ref{Fcomm5})}\leq \nonumber\\
 R_{31} + R_{21} + R_{32} + R_{3c} + R_{2c} \overset{(\ref{UB_5})}\leq  n_{2}  
\end{align*}
Similarly, for equation (\ref{OCTP5})
\begin{align*}
 d_{2} + d_{3} + c_{2} + 2b +  a_{2} + a_{3} &\overset{(\ref{OCTP2}),(\ref{TCOP6})}\leq \nonumber\\
R_{1c}^{'} + 2b +  a_{2} + a_{3} &\overset{(\ref{Fcomm1}),(\ref{Fcomm5})}\leq \nonumber\\
 R_{1c} + R_{2c} +  R_{23} + R_{13} &\overset{(\ref{UB_1})}\leq n_{3}
\end{align*}
After removing the occupied levels, the residual rate vector that still needs to be achieved is,
\begin{align}\label{OTCP8}
 \textbf{R}_\textbf{I}^{'''} &= (0,0,R^{''}_{1c} - d_{2} - d_{3},R^{''}_{21}-d_{2}-d_{3},0,0, \nonumber \\  
 &R^{''}_{31} - d_{3},R^{''}_{32}-d_{2},0) \nonumber \\ 
 &\triangleq (0,0,R^{'''}_{1c},R^{'''}_{21},,0,0,0,0,0)  
\end{align}   
 
 with number of levels at each node $(n^{'''}_{1},n^{'''}_{2},n^{'''}_{3})$ where,
 \begin{equation}\label{OCTP9}
 n^{'''}_{1} = n^{''}_{1} - 2d_{2} - 2d_{3}
 \end{equation}
 \begin{equation}\label{OCTP10}
 n^{'''}_{2} = n^{''}_{2} - 2d_{2} - 2d_{3}
 \end{equation}
 \begin{equation}\label{OCTP11}
 n^{'''}_{3} = \min\lbrace n^{''}_{3} -d_{2}-d_{3}, n^{'''}_{2}\rbrace
 \end{equation}
\item Full Private Scenario \\
 In this scenario, the transmission and reception strategy is similar to the cyclic communication scheme described in \cite{chaaban2011capacity}, where the users communicate in a rotational manner, either $1\longrightarrow 2 \longrightarrow 3 \longrightarrow 1$ or $1 \longrightarrow 3 \longrightarrow 2 \longrightarrow 1$. These two cycles can be represented by the following two equations 
 \begin{equation}\label{FP1}
  e_{1} = \min\lbrace R_{12}^{'''}, R_{23}^{'''}, R_{31}^{'''} \rbrace
  \end{equation} 
 \begin{equation}\label{FP2}
  e_{2} = \min\lbrace R_{13}^{'''}, R_{32}^{'''}, R_{21}^{'''} \rbrace
  \end{equation}  
Note that, as we always start with the maximum gain strategy, either $ e_1 $ or $ e_2 $ must be zero. Additionally,
from our selected rate tuple, $ e_1 = e_2 = 0 $. 
\end{enumerate}
\subsection{Minimum Gain Strategy}  
Finally, we want to achieve the remaining rate vector with the associated number of levels at each node in (\ref{OCTP9}),(\ref{OCTP10}) and (\ref{OCTP11}). It's obvious that the remaining rates in (\ref{OTCP8}) can not be achieved via the maximum gain strategy nor the intermediate gain one. Hence, we send each bit on a single level achieving  a gain of one bit per relay level.

 \par Considering our case: $R^{'''}_{1c} \neq 0 $ and $R^{'''}_{21} \neq 0$.
 Let the transmit vectors of users 1, 2 be $\boldsymbol{z}_{1c} = f_{1}$ , $\boldsymbol{z}_{21} = f_{2}$.
 User 1 transmits  $\boldsymbol{z}_{1c}$ on relay levels $\left\lbrace n^{'''}_{1} - f_{1} + 1,\cdots,n^{'''}_{1}\right\rbrace$.  
 User 2 sends $\boldsymbol{z}_{21}$ on relay levels $\lbrace n^{'''}_{2} - f_{2},\cdots,n^{'''}_{2}\rbrace$. Then the relay forwards $\boldsymbol{z}_{1c}$ on levels  $\left\lbrace 1,\cdots,f_{1}\right\rbrace$, and $\boldsymbol{z}_{12}$ on levels  $\{n^{'''}_{2}-f_{1}+1,\cdots,n^{'''}_{2}\}$. To guarantee the transmission of these bits, we should have $R_{1c}^{'''} + R_{21}^{'''} \leq n^{'''}_{1} $ and $R_{21}^{'''} \leq n^{'''}_{2} $ in uplink, and $R_{1c}^{'''} \leq n^{'''}_{3} $ in downlink. After combining we get,
    \begin{equation}\label{Min1}
    R_{21}^{'''} \leq n^{'''}_{2}  
    \end{equation}
    \begin{equation}\label{Min2}
    R_{21}^{'''} + R_{1c}^{'''} \leq n^{'''}_{1}
    \end{equation}
    \begin{equation}\label{Min3}
    R_{1c}^{'''} \leq n^{'''}_{3}
    \end{equation}
Starting from (\ref{Min1}), we should have $ R_{21}^{'''} \overset{(\ref{OCTP10}),(\ref{TCOP8})} \leq n_{2}^{'} -2c_{2} - 2d_{2} - 2d_{3} \overset{(\ref{Fcomm7}),(\ref{Max6})} = n_{2} -2c_{2} - 2d_{2} - 2d_{3} - 2b - a_{1} - a_{2} - a_{3}$, since $R_{21}^{'''} \overset{(\ref{Fcomm5}),(\ref{TCOP6}),(\ref{OTCP8})}= R_{21} - c_{2} - d_{2} - d_{3} - a_{1} $. Therefore, $R_{21} + c_{2} + d_{2} + d_{3} + 2b + a_{2} + a_{3} \overset{(\ref{OCTP2}),(\ref{OCTP3})} \leq R_{21} + c_{2} + R_{32}^{'} + R_{31}^{'} + 2b + a_{2} + a_{3} \overset{(\ref{TCOP3}),(\ref{Fcomm5})} \leq R_{21} + R_{3c} + R_{32}^{'} + R_{31}^{'} + b + a_{2} + a_{3} \overset{(\ref{Fcomm1}),(\ref{Max1})} \leq R_{21} + R_{3c} + R_{32} + R_{31} + R_{2c} \overset{(\ref{UB_5})} \leq n_{2}$.

In (\ref{Min2}), we should have $ R_{21}^{'''} + R_{1c}^{'''} \overset{(\ref{OCTP9}),(\ref{TCOP7})} \leq n_{1}^{'} -2c_{2} - 2d_{2} - 2d_{3} \overset{(\ref{Max5}),(\ref{Fcomm6})} = n_{1} -2c_{2} - 2d_{2} - 2d_{3} - 2b - a_{1} - a_{2} - a_{3}$, since $R_{21}^{'''} + R_{1c}^{'''} \overset{(\ref{Fcomm5}),(\ref{TCOP6}),(\ref{OTCP8})}= R_{21} + R_{1c} - b - 2c_{2} - 2d_{2} - 2d_{3} - a_{1}$. Therefore, $R_{21} + R_{1c} + b + a_{2} + a_{3} \overset{(\ref{Fcomm1}),(\ref{Max1})} \leq R_{21} + R_{1c} + R_{2c} + R_{13} + R_{23} \overset{(\ref{Ub_7})} \leq n_{1}$. 

In (\ref{Min3}), we should have $ R_{1c}^{'''} \overset{(\ref{OCTP11})} \leq \min\lbrace n^{''}_{3} -d_{2}-d_{3}, n^{'''}_{2}\rbrace$, since $R_{1c}^{'''} \overset{(\ref{Fcomm5}),(\ref{TCOP6}),(\ref{OTCP8})}= R_{1c} - d_{2} - d_{3} - c_{2} - b \overset{(\ref{UB_1})} \leq n_{3} -R_{2c}-R_{13}-R_{23} - d_{2} - d_{3} - c_{2} - b \overset{(\ref{Max1}),(\ref{Fcomm1})} \leq n_{3} -a_{2}-a_{3} - d_{2} - d_{3} - c_{2} - 2b \overset{(\ref{Max7}),(\ref{Fcomm8}),(\ref{TCOP9})} = n^{''}_{3} -d_{2}-d_{3}$. 

Also, from (\ref{Min3}),  we should have $R_{1c}^{'''} \leq n^{'''}_{2} $, as $R^{'''}_{1c} \overset{(\ref{Fcomm5}),(\ref{TCOP6}),(\ref{OTCP8})}= 
R_{1c} - b  - d_{2} - d_{3} - c_{2},$ and $n_2^{'''} \overset{(\ref{Max6}),(\ref{Fcomm7}), (\ref{TCOP8}), (\ref{OCTP10})} = n_{2} -2c_{2} - 2d_{2} - 2d_{3} - 2b - a_{1} - a_{2} - a_{3}$, so equation (\ref{Min3})
can be written as 
\begin{eqnarray}\label{Min4}
  R_{1c} + c_2 + d_2 + d_3 + b + a_1 + a_2 + a_3 \leq n_2 \label{Min5} \\ 
  R_{1c} + R_{31}^{'} + R_{32}^{'}+ R_{3c}^{'}  + b + a_2 + a_3 \overset{(\ref{OCTP3}),(\ref{OCTP2}),(\ref{TCOP6}),(\ref{TCOP3})}{\leq} n_2 \label{Min6}\\
  R_{1c} + R_{31} + R_{3c} + R_{32} + R_{12} \overset{(\ref{Fcomm5}),(\ref{Max1})}{\leq} n_2 \label{Min7}
\end{eqnarray} 

where (\ref{Min7}) is one of the terms in the minimum expression of (\ref{L_con}), and it can be simply shown that this term is less than the other one, that's why we take the minimum. However, in (\ref{Max0}), if we keep $ R_{2c}^{'} $ and $ R_{23}^{'} $  instead of $ R_{3c}^{'} $ and $ R_{32}^{'} $, respectively, and follow the same way, we get the following inequality
 \begin{equation}
   R_{1c} + R_{21} + R_{2c} + R_{23} + R_{13} \leq n_2 \label{Min8}
 \end{equation}  
which is the other term in the minimum expression of (\ref{L_con}), and again, it can be simply shown that this term is less than the other one. Hence, in order to serve the remaining bits, the condition in Lemma 1 should be satisfied.
 
\section{SINGLE SIDED GENIE UPPER BOUNDS FOR COMMON MESSAGES} \label{Upper_bound_common}
The relay channel can be represented as the combination of
multiple access channel i.e., uplink, and broadcast channel
i.e., downlink. In the traditional cut-set bounds
in \cite{cover2012elements}, the nodes are partitioned into two sets $S$
and $S^{c}$ which represent the transmitting and receiving relays,
respectively. As was mentioned in \cite{mokhtar2010deterministic}, if all nodes in $S^{c}$ fully cooperate and share all their side information, this cooperation is referred to as the two-sided genie aided bound. As was
shown in \cite{mokhtar2010deterministic}, applying this traditional cut-set bound to the
relay network produces loose bounds. Therefore, a tighter single-sided genie aided upper bound was developed in \cite{mokhtar2010deterministic}, where 
the notion of single sided genie comes from transferring information in only one direction by the genie.
Due to the existence of common messages in our network, we found that this bound is not tight at some regions. 
To show that, we consider one of the cuts around the relay. 
As shown in \cite{mokhtar2010deterministic}, we assume that the genie transfers only all data of node $ i $ to node $ j $ and $ k $, i.e., $(R_{ij},R_{ik},R_{ic})$. Also, it transfers all data of node $ j $ to node $ k $ only, i.e., $(R_{jk},R_{jc})$. 
Therefore, the data sent from node $ j $ to node $ i $ i.e. $(R_{ji},R_{jc})$ is not known at node $ i $ a priori. 
Also, the data sent from node $ k $ to both nodes $ i $ and $ j $ i.e. $(R_{kj},R_{ki},R_{kc})$ is not known at both of them. 
As shown here, $R_{jc}$ represents a bottleneck because we assumed that the genie transfers messages from node $ j $ to node $ k $ only. However, it is not known at node $ i $. This results in a looser inequality,
  \begin{equation} \label{gen}
 R_{kj} + R_{ki} + R_{kc} + R_{ji} + R_{jc} \leq \max(n_{i},n_{j},n_{k})
  \end{equation}  
In the above equation, we did not exploit the common messages information transferred by the genie from user $ j $ to user $ k $.  As a result, we believe that this upper bound is not tight in existence of common messages.
\section{Discussion}
\label{disc}
In Section \ref{Upper_bound_common}, we argued that the single-sided genie bounds and the cut-set bounds are not sufficient to characterize the capacity region of the networks with common messages. As the common message will be known at one node only, however, we should send it to the other one, and this will lead to a looser inequality (\ref{gen}). As a result, for the non-achievable rate tuples, we found that we need extra relay levels to send the common message as illustrated in example 2. Our results reveals that, when $n_{1} = n_{2}$, the upper bound is achievable. Additionally, it is important to mention that we only need the extra condition of Lemma 1 if and only if $ d_3 \neq 0 $ in (\ref{OCTP3}), otherwise, any rate tuple is achievable using our greedy scheme. 
We illustrate our work through two examples. The rate tuple in the first example is achievable while the other one violates (11) and hence, can not be achieved. Consider a reciprocal network with channel
gains $(n_{1}, n_{2}, n_{3}) = (6, 5, 4)$ and the rate
tuple $ R = (R_{12},R_{13},R_{1c},R_{21},R_{23},R_{2c},R_{31},R_{32},R_{3c}) $
\par \textit{Example 1:} $R =(1, 1, 1, 1, 0, 2, 0, 0, 2)$ which satisfies the upper bounds in Theorem 1,
and satisfies condition from Lemma 1. This rate tuple is achievable with the GOS.

\begin{figure}
  \centering
\includegraphics[width=0.9\linewidth]{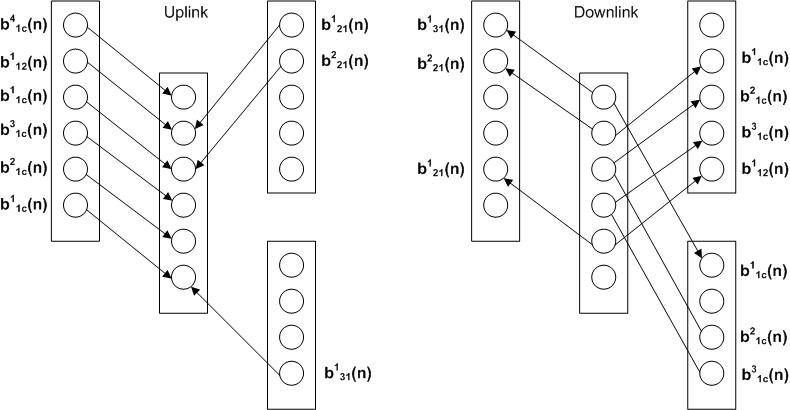}
\caption{Example 2      \label{Example2}}
\end{figure}

 \textit{Example 2:} $R =(1, 0, 4, 2, 0, 0, 1, 0, 0)$ which satisfies the upper bounds in Theorem 1,
but violates condition from Lemma 1. We found here that we need to send one of the common messages on two levels. As shown in Fig. \ref{Example2}, the problem appeared only in the downlink because the common message should be received at two nodes in the downlink while we send it once in the uplink. In addition, it was counted only one time in the upper bound equations in Section II.

It is worth mentioning that we studied the weighted sum Degrees of Freedom (DoF) of the MIMO Y channel with common messages and private messages in \cite{salahglo2017}. However, as the DoF can be considered as the total number of received interference-free streams, and as each common message should be successfully decoded at two different receivers, we can scale the DoF of each common message by a factor of 2. As a result, we were able to achieve the upper bound on the total DoF of that network and got very insightful results. However, in the deterministic sense, since we are studying the entire region of the network, it will not be possible to weight the common message by a factor of 2 because we are dealing with a rate tuple, not a metric. Therefore, it makes sense to have some tuples that can not be achieved because the common message is counted only once in the upper bound equations.

\begin{figure}
  \centering
\includegraphics[width=0.9\linewidth]{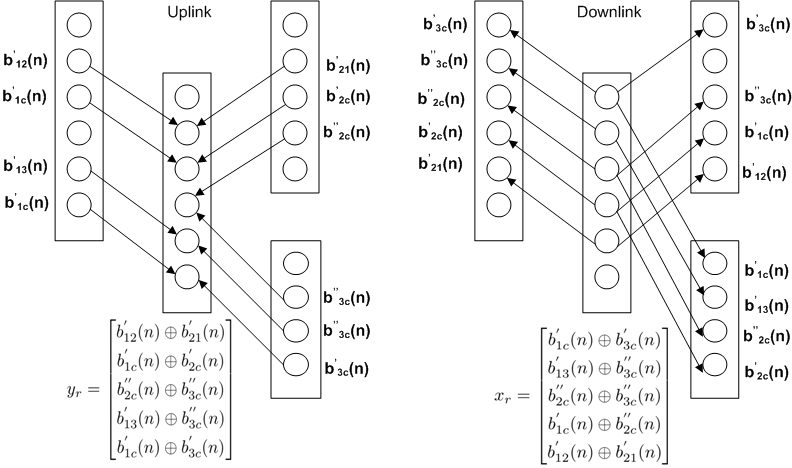}
\caption{Example 1     \label{Example1}}
\end{figure}

\section{CONCLUSIONS}
\label{conc}
In this work, the deterministic Y-channel with private and common messages was considered. All the users communicate with each other via a relay. Our work results in defining an outer bounds based on the notion of single sided genie. We showed that this outer bound is not tight at some regions due to the existence of common messages. We characterize the achievable region using the Gain Ordering Scheme which depends mainly on starting with strategy that achieves a maximum gain and ending up with the one of minimum gain. We are convinced that this scheme is optimal and the remaining rate tuples is not achievable. Numerical examples are explained to illustrate the operation of our scheme. As a future work, we believe that a novel outer bound which is tighter than the single sided-genie can be considered. In addition, this work paves the way for calculating the capacity of gaussian channels with private messages and common messages.
\bibliographystyle
{IEEEtran}
\bibliography{IEEEabrv,ref}

\end{document}